\documentstyle[aps,prbbib,twocolumn,epsf]{revtex}

\begin{document}
\draft
\title{{\it ab initio} modeling of open systems: charge transfer, 
electron conduction, and molecular switching of a $C_{60}$ device}

\author{Jeremy Taylor$^1$, Hong Guo$^1$, and Jian Wang$^2$}

\address{1. Center for the Physics of Materials and Department 
of Physics, McGill University, Montreal, PQ, Canada H3A 2T8.\\
2. Department of Physics, The University of Hong Kong, 
Pokfulam Road, Hong Kong, China.\\
}
\maketitle

\begin{abstract}

We present an {\it ab initio} analysis of electron conduction through a
$C_{60}$ molecular device. Charge transfer from the device electrodes to 
the molecular region is found to play a crucial role in aligning the lowest 
unoccupied molecular orbital (LUMO) of the $C_{60}$ to the Fermi level of 
the electrodes.  This alignment induces a substantial device conductance of 
$\sim 2.2 \times (2e^2/h)$. A gate potential can inhibit charge transfer 
and introduce a conductance gap near $E_F$, changing the current-voltage 
characteristics from metallic to semi-conducting, thereby producing a field
effect molecular current switch.

\end{abstract}

\pacs{72.80.Rj,73.61.Wp,73.23.Ad}

Understanding electron conduction in atomic and molecular scale nano-devices is
an extremely active research topic at present. In several recent 
experiments\cite{gimzewski,eigler,tans,dekker,reed,datta1}, electric current 
flowing through truly atomic and molecular scale conductors, as small as
a single atom, have been measured. The current voltage (I-V) characteristics 
of these molecular devices have profound potential for device application, 
including high nonlinearity, negative differential resistance, and 
electro-mechanic current switching. However, a thorough understanding of 
electron transport mechanisms at this scale, which will play a crucial role 
in designing operation principles for future nanoelectronics, has not yet 
been achieved. To provide insight into electron conduction mechanisms at 
the atomic and molecular scale, we have investigated the quantum transport 
properties of a $C_{60}$ based molecular device using a first principles 
theoretical method.

Experimentally, the $C_{60}$ molecular electro-mechanical 
amplifier\cite{gimzewski} demonstrated that electric current flowing through 
a $C_{60}$ molecule could be amplified by as large as $100$ times when 
the molecule was {\it mechanically} deformed. This phenomenon was attributed 
to an increase of the tunneling density of states at the Fermi level induced 
by the mechanical deformation\cite{gimzewski,joachim}. While this conduction 
mechanism is very interesting, a natural question is whether or not 
there are other, perhaps simpler, physical mechanisms which can be exploited 
for molecular scale electron conduction ? Our analysis suggests a unique and 
new conduction mechanism based on charge transfer doping (CTD): because of its 
high electro-negativity, a well contacted $C_{60}$ gains charge from CTD 
thereby aligning its LUMO state to the Fermi level of the electrodes resulting 
in a substantial device conductance without the need of mechanical 
deformation. CTD was seen to have {\it reduced} the conductance of a short 
carbon chain\cite{lang1}, but for the more complex $C_{60}$ molecule, it does 
the opposite by drastically {\it increasing} conduction. Current switching 
mechanisms are another important yet unsettled question of nanoelectronics.
Our analysis further suggests a new principle for molecular switching: using 
proper electro-negative molecules and metallic electrodes, current can be 
switched on and off by controlling the amount of charge that is transferred 
to the molecular region.


The $C_{60}$ device we study is illustrated schematically in Fig.(1a) where 
a $C_{60}$ molecule is electrically {\it well} contacted by two atomic scale 
Al metallic electrodes which extend to reservoirs far away 
where bias voltages $V_{l,r}$ are applied to the left ($l$) and right ($r$) 
electrodes, respectively. An additional gate voltage $V_g$ may also be 
applied to a metallic gate capacitively coupled to the molecule. Due to the 
large number of atoms and complications such as the gate, the bias, the 
atomic electrodes (as opposed to jellium electrodes), and the presence of 
localized states, existing {\it ab initio} methods for analyzing quantum 
transport\cite{lang,wan,ihm} cannot be applied. We have therefore developed 
a new approach which combines non-equilibrium Green's function 
theory\cite{jauho,jian1} with pseudo-potential real space {\it ab initio} 
simulation techniques\cite{ordejon}. It is worth noting that conventional 
density functional theory (DFT) methods solve problems for either 
finite systems such as an isolated molecule, or periodic systems consisting 
of super-cells. In contrast, central to our problem of predicting quantum 
transport properties is the ability to deal with systems having {\it open} 
boundaries provided by long electrodes which maintain different chemical 
potential due to external bias. In other words, a typical device geometry, 
{\it e.g.} Fig.(1a), is neither isolated nor periodic. This situation is 
most conveniently handled by nonequilibrium Green's function 
theory\cite{jauho,jian1}.

Briefly, our technique is outlined as follows\cite{jeremy1}. We divide the 
long device system into three regions\cite{wan}: the left and right 
electrodes, and the scattering region (Fig.(1a)). The scattering region 
actually includes a portion of the semi-infinite electrodes\cite{wan}. 
After self-consistency is reached, this arrangement ensures good electric 
contact between the electrodes and the molecule, establishes a common Fermi 
level for the system, and ensures charge neutrality at equilibrium. When the 
scattering region is large enough\cite{wan,jian2}, the Kohn-Sham (KS) potential
outside the scattering region is well screened and therefore well approximated 
by a perfect ``bulk'' electrode environment which we obtain by a separate 
calculation\cite{jian2}. By setting the KS potential outside the scattering 
region to the bulk value and matching the electrostatic potential at the 
boundary (including the gate), the {\em infinite} open boundary problem 
is reduced to the calculation of charge density inside the {\em finite} 
scattering region with the electrodes' contribution accounted for by 
appropriate self-energies\cite{jauho,jian1,datta_book}. 

To perform our self-consistent DFT analysis, we calculate the charge density 
from the density matrix which is related to nonequilibrium Green's 
function\cite{jauho,jian1,datta_book} operator,
\begin{equation}
\hat{\rho}  = i \int_{-\infty}^{\infty} dE \, {\bf G}^<(E) =
i \int_{-\infty}^{\infty} dE \, {\bf G}^R {\bf \Sigma}^{<} {\bf G}^A \\
\label{rho1}
\end{equation}
where\cite{jauho,jian1,datta_book}:
\begin{equation}
{\bf \Sigma}^{<} = \left[ 
	\begin{array}{ccc}
	f_l(E;\mu_l) {\bf \Sigma}^E_{l,l} & 0 & 0 \\
	0 & 0 & 0 \\
	0 & 0 & f_r(E;\mu_r) {\bf \Sigma}^E_{r,r} \end{array} \right]
\end{equation}
and $f_{l,r}(E;\mu_{l,r})$ are the distribution functions deep in the left 
and right electrodes with chemical potentials $\mu_{l,r}$, respectively. The 
reservoirs are assumed to be at equilibrium, therefore $f_{l,r} \approx 
\Theta(E-\mu_{l,r})$ at low temperatures where $\Theta$ is the step function. 
The retarded (and advanced) Green's functions ${\bf G}^R$ (${\bf G}^A$) of 
the system are then calculated in standard 
fashion\cite{datta_book,lambert,guo1} by direct matrix inversion where 
self-energies due to coupling to electrodes 
(${\bf \Sigma}^E_{l,l},{\bf \Sigma}^E_{r,r}$) are calculated by extending 
the method of Ref.\onlinecite{lambert}. Because there are many bands in our 
atomic electrodes (as compared to jellium electrodes\cite{lang,wan}), the 
energy integration of Eq. (\ref{rho1}) must be done with care so that global 
charge neutrality is maintained at equilibrium. With the charge density, we 
can evaluate the effective device potential $V_{\rm eff} [\rho({\bf r})]$ 
which consists of Hartree, exchange-correlation, atomic core, and any other 
external potentials\cite{jeremy1}. Using the boundary condition on the 
effective potential discussed above, $V_{\rm eff}({\bf r})$ is defined 
everywhere in space. It is therefore straightforward to construct the 
Hamiltonian matrix using an $s,\ p$ Fireball atomic orbital basis 
set~\cite{ordejon}. We use standard norm-conserving pseduopotentials to 
describe the atomic cores~\cite{hsc}. The DFT iteration is repeated until 
self-consistency, and current is calculated by integrating conductance $G(E)$ 
over energy\cite{jauho,jian1,datta_book}, with
$G(E)= 4G_o \mbox{ Tr}[Im({\bf \Sigma}_{ll}^E){\bf G}^R Im({\bf \Sigma}_{rr}^E)
{\bf G}^A]$ where $ G_o = \frac{2e^2}{h} $.

Fig.(1b) shows the equilibrium charge distribution along the middle 
cross-section of the device\cite{foot3} at zero bias and gate voltages. 
The charge density in the electrodes is affected by the $C_{60}$
but this effect is well screened away from the molecule. Indeed, the charge
density contours match perfectly at the connections between the electrodes
and the scattering region\cite{foot3}. The electronic structure  
of an isolated $C_{60}$ molecule is well known\cite{book} and our 
{\it ab initio} calculation gives a HOMO-LUMO gap of $1.77$eV, 
in excellent agreement with previous literature\cite{book1}. The undoped
$C_{60}$ {\it molecular solid} is a semiconductor\cite{book}. Since $C_{60}$ is 
electro-negative, one can dope the solid and fill the LUMO state with up 
to 6 electrons~\cite{book}. Importantly, for our device the {\it open} 
metallic electrodes provide natural doping through CTD. As a result,
we found that, in equilibrium, three extra electrons flow into the 
molecular region. This is a substantial charge transfer in order for the 
molecule to equilibrate with the electrodes so that a common Fermi level is 
established.  Experimentally, it is found that the $C_{60}$ solid
conducts best when it is doped with three electrons per $C_{60}$\cite{book}.
This is because doping three electrons amounts to half filling the LUMO. 
Our {\it ab initio} analysis predicts, for the $C_{60}$ device, that the 
$C_{60}$ LUMO state is half-filled and we find an equilibrium conductance 
of $G(E_F)=2.2 G_o$ at zero temperature. This value is very significant: 
without charge transfer we would expect a much smaller conductance due to 
the filled HOMO state and the substantial HOMO-LUMO gap of the isolated 
$C_{60}$.


In Fig.(2), we plot the equilibrium transmission eigenvalues of the device
as a function of electron energy at two different gate voltages. At $V_g=0$,
there are three transmission eigenvectors which contribute substantially at 
$E_F$ (the vertical line) to the equilibrium conductance $G = 2.2 G_o $.
However, a gate voltage shifts the states near $E_F$ and changes the 
transmission channels significantly.  This is shown in Fig.(2b) where we 
see a conductance gap at $E_F$. Therefore, a gate potential can change electron 
conduction through this device significantly, producing field induced 
molecular switching. We can understand the transport properties by projecting 
the scattering states of the device on to the molecular orbitals of an 
isolated $C_{60}$. We found that scattering states near $E_F$ are over 
90\% HOMO or LUMO in character, indicating that the $C_{60}$ retains its 
isolated electronic structure and that other $C_{60}$ orbitals are less 
important for conduction. In addition, we can classify each scattering state 
as being majority HOMO or majority LUMO. We find that majority HOMO states 
are almost pure HOMO ($>90$\%); while majority LUMO states are a HOMO-LUMO 
mixture ($\sim$30:70). Specifically, the transmission eigenvectors of 
Fig.(2a) are all LUMO-like, indicating that there are three electrons half 
filling the LUMO state and contributing three conduction channels at 
equilibrium. In other words, the LUMO state is naturally aligned with $E_F$ 
due to charge transfer. On the other hand, a negative gate voltage can inhibit 
charge transfer so that $E_F$ lies between the HOMO and LUMO states of the 
molecule. Indeed, we found that the scattering state corresponding to the left 
peak in Fig.(2b) is HOMO-like, and that of the right peak is LUMO-like. 
In the absence of charge transfer, one would expect transmission properties
similar to those in Fig. (2b).


The above physical picture is further demonstrated in Fig. (3) which shows
the equilibrium conductance $G$, the three individual transmission eigenvalues 
$T_i$ ($i=1,2,3$), and the number of transferred charge $Q$ (inset) as 
functions of gate voltage $V_g$ at $E_F$. $G$ has a step-like behavior: 
each ``step'' is due to the depletion of a LUMO-like state. Only after charge 
transfer is completely inhibited do we obtain the conductance gap illustrated 
in Fig.(2b) where $G$ becomes very small. Our result therefore indicates that 
charge transfer plays a crucial role and changes the physical picture of 
electron conduction in this device qualitatively. The behavior of $G$ also has 
a clear correspondence with $T_i$ (the dotted lines with circles): as $V_g$ 
is scanned toward more negative values, $T_i$'s are removed one by one.

Charge transfer also has very important implications for the I-V curves, 
shown in Fig.(4). The current is plotted as a function of right bias $V_r$ 
fixing $V_l=0$. When $V_g=0$ the I-V curve shows clear metallic behavior.
When $V_g\neq 0$, the $C_{60}$ device can change from a metal (Fig. (2a)) 
to a semiconductor (Fig. (2b)) with a conductance gap near zero bias. 
Therefore the gate potential can ``switch'' off the current, reflecting the
transition of transmission eigenvalues from Fig.(2a) to (2b). This corresponds
to an amplification factor of $\sim 20$. The value of gate potential is 
non-universal as it depends on the shape of the gate and dielectric medium 
surrounding the device. What is essential is to produce enough electric 
field lines inside the molecular region so that the electrode-doped LUMO 
electrons are depleted from the molecular junction.  For our system, 
a shift of $\sim 0.1$eV is enough to generate switching. This is however a 
large shift and it suggests that a sharp-shaped gate is perhaps necessary in 
an experimental setup, otherwise most field lines will be screened by the 
metallic electrodes. The predicted current is in the range of $ \sim 10 \mu A$ 
at small bias voltages ($\sim 50$mV). This is somewhat larger than the 
$ \sim 4 \mu A $ measured for the electro-mechanic 
amplifier\cite{gimzewski,foot5}.


Our results also have important implications and provide a benchmark for 
semi-empirical and other theories, such as those based on parameterized tight 
binding models. The chemical potential difference between an isolated Al 
electrode and an isolated $C_{60}$ molecule
is $\Delta\equiv E_F^{electrode}-E_F^{C_{60}} \sim 0.19 a.u.$. This provides 
a lot of uncertainty as how one should align the molecular orbitals to the 
Fermi level of the electrodes in a non-self-consistent calculation. 
Specifically\cite{foot4}, calculating $G(E_F^{electrode})$ without alignment 
gives $1.17G_o$, much less than the correct result. Shifting levels of 
$C_{60}$ by adjusting a chemical potential so that global charge neutrality 
is obtained, we get $1.7G_o$. Shifting levels so that three extra charges are 
inside $C_{60}$ region, the result becomes $2.0 G_o$. Because we cannot know 
charge transfer in a non-self consistent analysis, the result suggests that 
requiring global charge neutrality in a semi-empirical calculation is the next 
best thing to do at equilibrium. In this regard, we note that there has been 
some practice for requiring local charge neutrality on each 
atom\cite{pernas,brandbyge} in semi-empirical calculations. Another 
interesting result we found is that there are many localized states in the 
molecular junction. At equilibrium, we find $96$ bound states by integrating 
the density of states (Eq.(\ref{rho1})) from $-\infty$ to the propagating 
threshold of the electrodes. These localized states play an important role in 
establishing the effective potential $V_{\rm eff}$ and must be included in 
the analysis.

In summary, our results suggest that charge transfer doping induces a 
substantial conductance in a well contacted molecular $C_{60}$ device. 
Essentially, CTD aligns the LUMO states of the $C_{60}$ with the Fermi 
level of the electrodes, opening up 3 conductance channels and producing 
metallic I-V characteristics at equilibrium. A field effect provided by a 
gate potential can switch off the current by inhibiting charge transfer. 
This is very interesting from a device operation point of view and provides 
a new mechanism for molecular scale current switches, complementing the 
electro-mechanical operation principle explored in previous 
work\cite{gimzewski}. The charge transfer induced conduction discussed here 
should be a more general mechanism: using proper electrodes which provide 
natural doping to electro-negative molecular systems with a filled HOMO state 
and substantial HOMO-LUMO gap, an increase in conduction is expected. Since 
one expects charge transfer to be a generic feature of molecular devices, 
this effect should be exploited further for applications of molecular scale 
current switches.

{\bf Acknowledgments:} We gratefully acknowledge financial support
from NSERC of Canada and FCAR of Quebec (H.G); RGC grant (HKU 7215/99P) from
the Hong Kong SAR (J.W.).  J.T gratefully acknowledge financial support
from NSERC through a PG Fellowship.

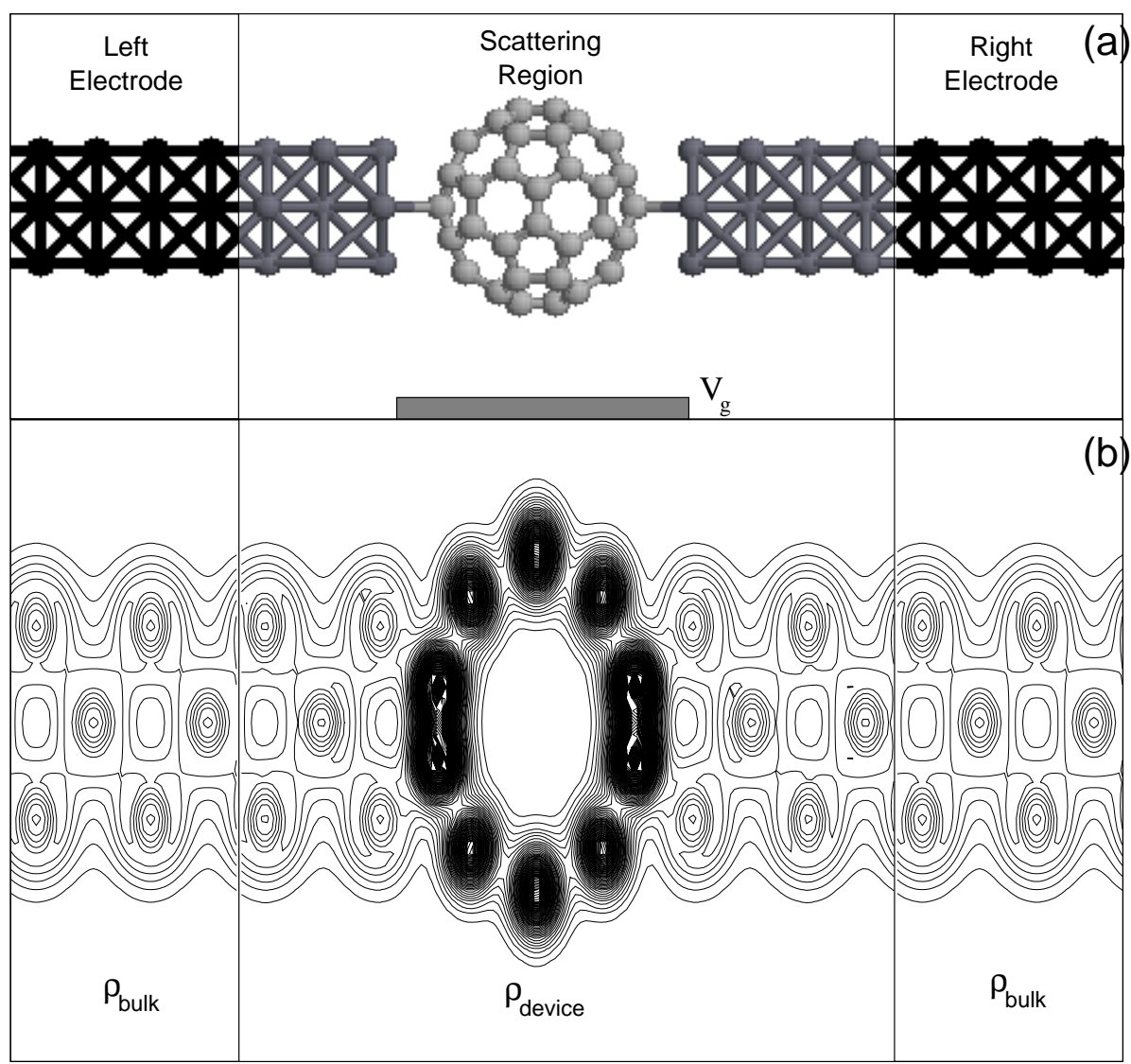
\begin{figure}
\caption{
(a). Schematic plot of the $C_{60}$ molecular device. (b). Contour plot of 
the equilibrium charge density.  Notice the perfect match across the 
boundaries between the scattering region and the electrodes.
}
\end{figure} 

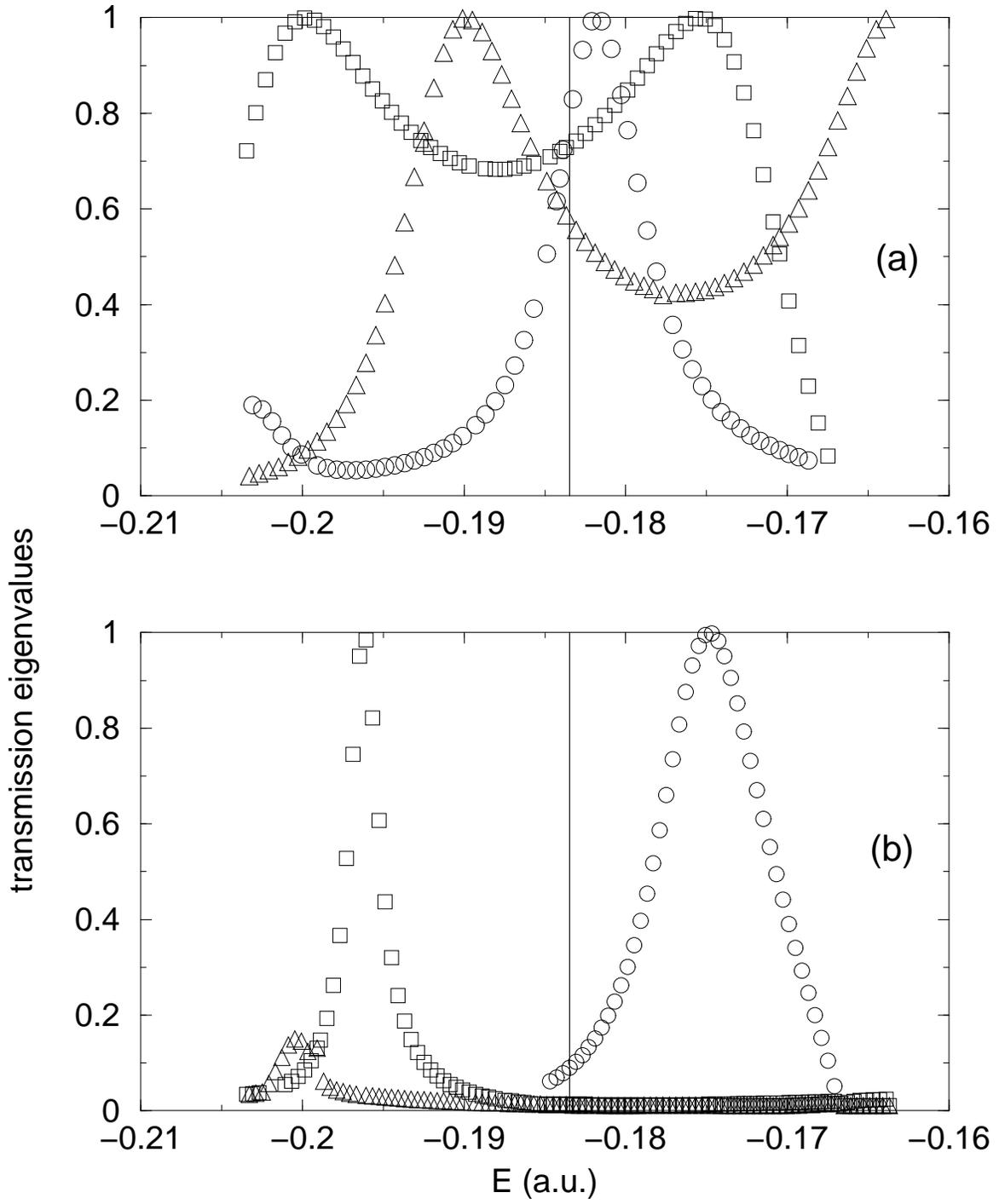
\begin{figure}
\caption{
Transmission eigenvalues as a function of electron energy. (a) For $V_g=0$ 
where there are three transmission eigenvalues. (b) For $V_g=1$a.u. 
Vertical line shows the Fermi level of the system.
} 
\end{figure}

\begin{figure}
\caption{
Solid squares: equilibrium conductance $G(E_F)$ as a function of gate 
voltage $V_g$. Vertical dotted lines indicate, approximately, integer number 
of transferred charges $Q$. Open circles: the three transmission eigenvalues 
as a function of $V_g$ at $E_F$. Inset: the transferred charge as a function 
of gate voltage.
} 
\end{figure}

\begin{figure}
\caption{
I-V curves of the system showing the switching between metallic (squares)
and ``semiconducting'' (circles) behavior.
} 
\end{figure}

\begin{thebibliography}{99}
\bibitem{foot3}
For simplicity we have fixed the atomic positions. The electrode-molecule
distance is kept at $1.06$\AA ~for the results presented here. We have
confirmed that there is no qualitative change to the results when this
distance is varied so long as bonding between electrodes and the molecule
exists. An Al electrode is represented by a slab of Al oriented along the 
(100) plane with eighteen atoms per unit cell. We have confirmed that doubling 
the electrode length (the portion included inside the scattering region) gives 
essentially the same result. The matching of the charge density contours 
(Fig.(1b)) gives a strong confirmation of the numerical accuracy as the 
boundary condition on the potential has generated a perfect match of the 
charge density.
\bibitem{book1}
Average gap value from the literature is $1.8$eV. See, for example, p458 of 
Ref.\onlinecite{book}.
\bibitem{foot5}
We do not expect an exact agreement because of the different physical
mechanisms for conduction. In addition, the $C_{60}$ in our device is very
well contacted which tends to give a larger current.
\bibitem{foot4}
We have carried out the non-self-consistent calculation using the Harris
functional.
\end{thebibliography}
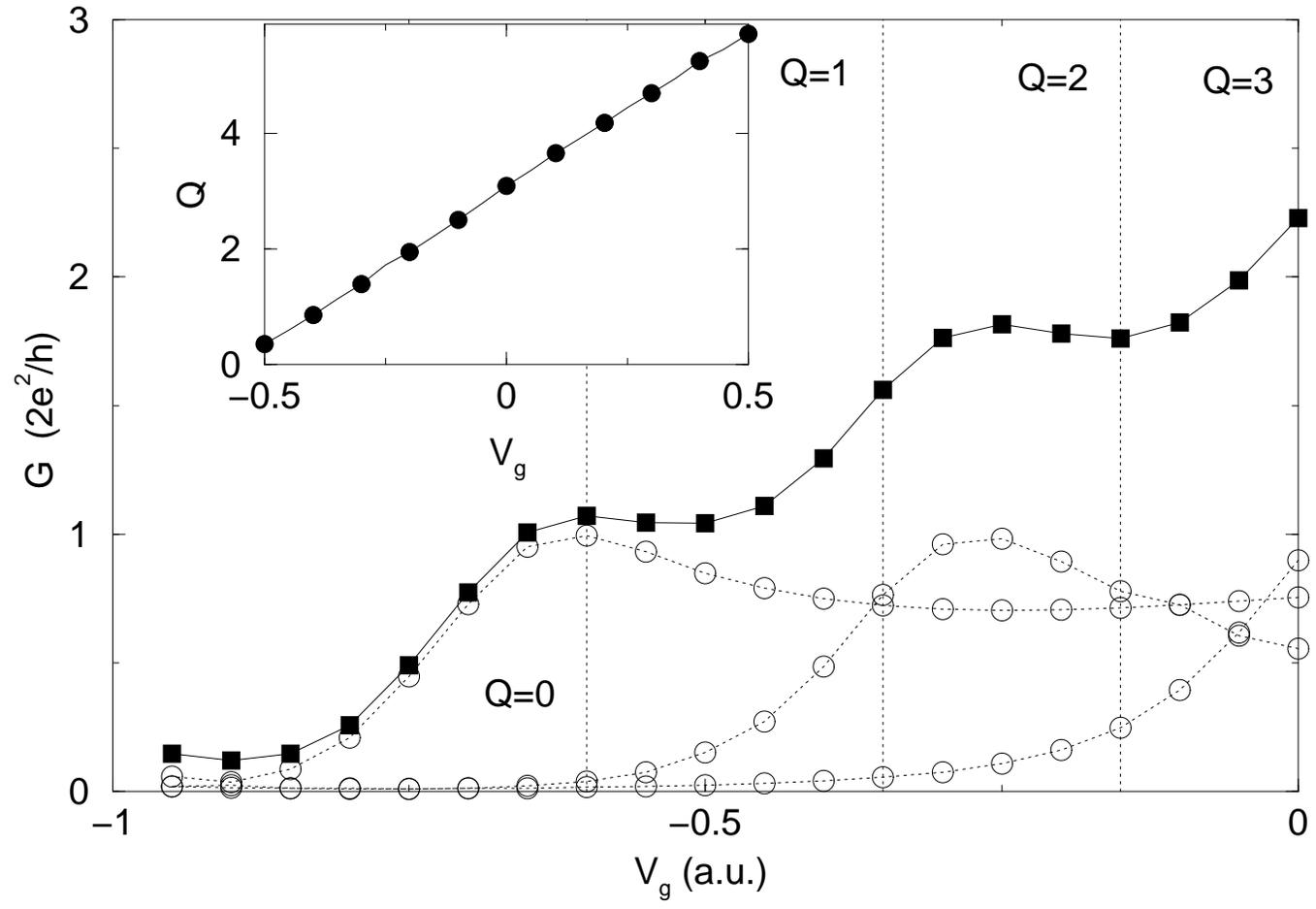
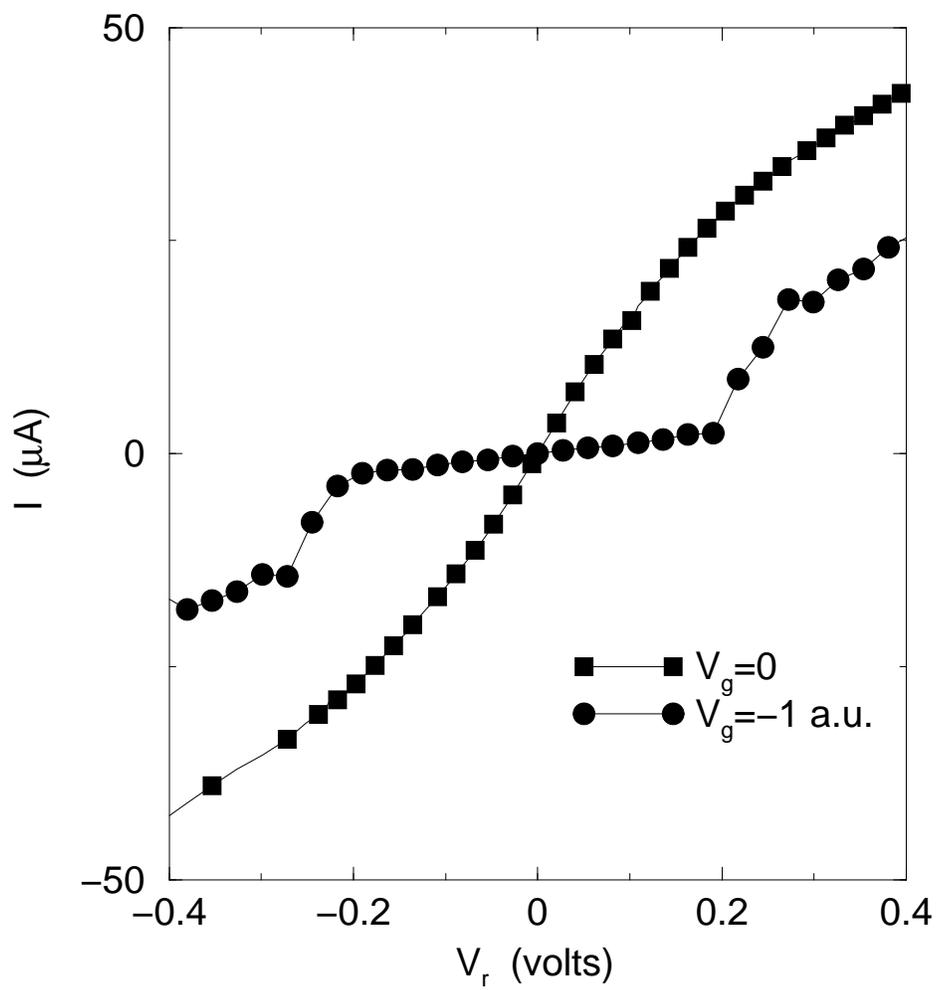
\end{document}